# Enriched Performance on Wireless Sensor Network using Fuzzy based Clustering Technique

A.M Nirmala
Assistant Professor
Dept. of Computer Science
Muthayammal Arts &
Science College, Rasipuram,
Tamilnadu, India.

P. Subramaniam
Assistant Professor
Dept. of Computer Science
Muthayammal Arts &
Science College, Rasipuram,
Tamilnadu, India

A. Anusha Priya
Assistant Professor
Dept. of Computer Science
Muthayammal Arts &
Science College, Rasipuram,
Tamilnadu, India

M. Ravi
Assistant Professor
Dept. of Computer Science
Muthayammal Arts &
Science College, Rasipuram,
Tamilnadu, India

*Abstract:* The wireless sensor networks combines sensing, computation, and communication into a single small device. These devices depend on battery power and may be placed in hostile environments replacing them becomes a tedious task. Thus improving the energy of these networks becomes important. Clustering in wireless sensor network looks several challenges such as selection of an optimal group of sensor nodes as cluster, optimum selection of cluster head, energy balanced optimal strategy for rotating the role of cluster head in a cluster, maintaining intra and inter cluster connectivity and optimal data routing in the network.

      In this paper, we study a protocol supporting an energy efficient clustering, cluster head selection and data routing method to extend the lifetime of sensor network. Simulation results demonstrate that the proposed protocol prolongs network lifetime due to the use of efficient clustering, cluster head selection and data routing. The results of simulation show that at the end of some certain part of running the EECS and Fuzzy based clustering algorithm increases the number of alive nodes comparing with the LEACH and HEED methods and this can lead to an increase in sensor network lifetime. By using the EECS method the total number of messages received at base station is increased when compared with LEACH and HEED methods. The Fuzzy based clustering method compared with the K-Means Clustering by means of iteration count and time taken to die first node in wireless sensor network, as the result shows that the fuzzy based clustering method perform well than kmeans clustering methods.

## I. INTRODUCTION

### A. Wireless Sensor Network (WSN)

Wireless sensor network is a collection of sensor nodes interconnected by wireless Communication channels. Each Sensor node is a small device that can collect data from its surrounding area, carry out simple computations, and communicate with other Sensors or with the base station (BS).

Recent years have observed an increasing interest in using wireless sensor networks (WSNs) in many applications, including environmental monitoring and military field surveillance. In these applications, small sensors are deployed and left unattended to continuously report parameters such as temperature, pressure, humidity, light, and chemical activity. Reports transmitted by these sensors are collected by observers (e.g., base stations).The dense deployment and unattended nature of WSNs makes it quite difficult to recharge node batteries [2], [4].

The concept of wireless sensor networks is based on a simple equation:

$$\text{Sensing} + \text{CPU} + \text{Radio} = \text{Thousands of potential applications}$$

### B. Clustering

Clustering is a separation of data into groups of similar objects. Each group called cluster consists of objects that are similar between themselves and dissimilar to objects of other groups.

#### 1. *Clustering in wireless sensor network* [5]

In clustering, the sensor nodes are partitioned into different clusters. Each cluster is managed by a node denoted as cluster head (CH) and other nodes are referred as cluster nodes. Cluster nodes do not communicate directly with the sink node. They have to pass the collected data to the cluster head. Cluster head will aggregate the data, received from cluster nodes and transmits it to the base station. Thus minimizes the energy consumption and number of messages communicated to base station. Also number of active nodes in communication is reduced. Ultimate result of clustering the sensor nodes is prolonged network lifetime.

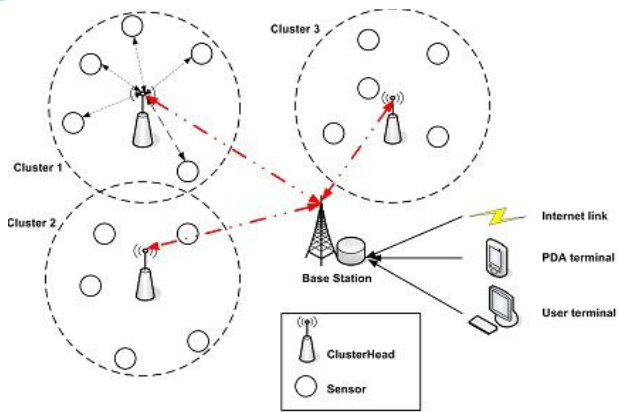

Figure 1. Clustering sensor nodes

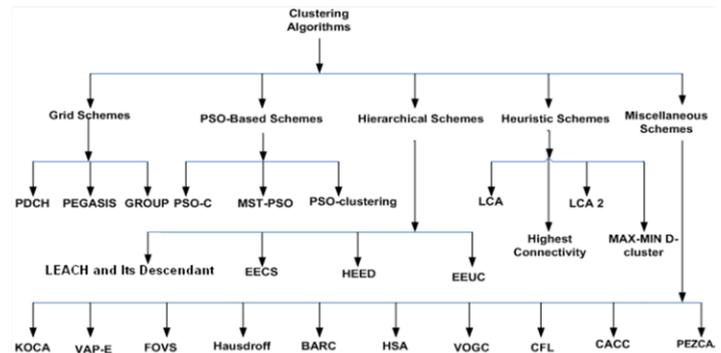

Figure 2. Taxonomy of clustering algorithms in WSNs

*Sensor Node:* It is the essential component of wireless sensor network. It has the capability of sensing, processing, routing, etc.

*Cluster Head:* The Cluster head (CH) is considered as a representative for that specific cluster and it is responsible for different activities carried out in the cluster, such as data aggregation, data transmission to base station, scheduling in the cluster, etc.

*Base Station:* Base station is considered as a main data collection node for the complete sensor network. It is the bridge (via communication link) between the sensor network and the end user. Normally this node is reflected as a node with no power constraints.

*Cluster:* It is the organizational unit of the network, created to streamline the communication in the sensor network.

*Advantages of Clustering*

- Reducing amount of nodes taking part in transmission
- Useful Energy consumption
- Scalability for large number of nodes
- Reduces communication overhead
- Efficient use of resources in WSNs

This paper is organized as follows. Section II presents an overview of Clustering algorithm in WSN. Section III describes Clustering techniques and its method. Section IV describes performance of Experimental analysis and discussion. Section V presents conclusion and future work.

## II. CLUSTERING ALGORITHMS IN WSN

Fig.2 shows the taxonomy of clustering algorithms in WSNs

### A. Low Energy Adaptive Clustering Hierarchy (LEACH)

Low Energy Adaptive Clustering Hierarchy is designed for sensor networks here an end-user wants to remotely monitor the environment. In such a situation, the data from the individual nodes must be sent to a central base station, often located far from the sensor network, through which the end-user can access the data. There are several desirable properties for protocols on these networks:

- Use 100's - 1000's of nodes
- Maximize system lifetime
- Maximize network coverage
- Use uniform, battery-operated nodes.

Conventional network protocols, such as direct transmission, minimum transmission energy, multi-hop routing and clustering all have drawbacks that don't allow them to achieve all the desirable properties. LEACH includes distributed cluster formation, local processing to reduce global communication, and randomized rotation of the cluster-heads. Together, these features allow LEACH to achieve the desired properties. Initial simulations show that LEACH is an energy-efficient protocol that extends system lifetime.

### B. Hybrid Energy-Efficient Distributed Clustering (HEED)

Nodes in LEACH independently decide to become cluster heads. While this approach requires no communication overhead, it has the drawback of not guaranteeing that the cluster head nodes are well distributed throughout the network. While the LEACH-C protocol solves this problem, it is a centralized approach that cannot scale to very large numbers of sensors. Many papers have proposed clustering algorithms that create more uniform clusters at the expense of overhead in cluster formation. One approach that uses a distributed algorithm that can converge quickly and has been shown to have low overhead is called HEED [10]. HEED uses an iterative cluster formation algorithm, where sensors assign themselves a "cluster head probability" that is a function of their residual energy and a "communication cost" that is a function of neighbor proximity. Using the cluster head probability, sensors decide whether or not to advertise that they are a candidate cluster head for this iteration. Based on these advertisement messages, each sensor selects the candidate cluster head with the lowest "communication cost" as its tentative cluster head. This procedure iterates, with each sensor increasing its cluster head probability at each iteration until the cluster head probability is one and the sensor declares itself a "final cluster head" for this round. The advantages of HEED are that node s only require local (neighborhood) information to form the clusters, the algorithm terminates in O (1) iterations, the algorithm guarantees that every sensors is part of just one cluster, and the cluster heads are well-distributed.

### C. *EECS: Energy Efficient Clustering Schemes* [6]

We introduce an algorithm in which cluster formation is different from LEACH protocol. In LEACH protocol cluster formation takes place on the basis of a minimum distance of nodes to their corresponding cluster head. In EECS [1], dynamic sizing of clusters takes place which is based on cluster distance from the base station. The results are an algorithm that addresses the problem that clusters at a greater distance from the sink requires more energy for transmission than those that are closer. Ultimately it provides equal distribution of energy in the networks, resulting in network lifetime. Thus main advantage of this algorithm is the full connectivity can be achieved for a longer duration. So we can say it provides reliable sensing capabilities at a larger range of networks for a longer period of time. It provides a 35 percent improvement in network life time over LEACH algorithm.

### III. CLUSTERING TECHNIQUES ON WIRELESS SENSOR NETWORK

### A. *Clustering and Cluster Head Selection [3] using LEACH.*

The operation of LEACH is broken up into rounds, where each round begins with a setup phase, when the clusters are organized, followed by a steady state phase, when data transfers to the base station occur. In order to minimize overhead, the steady-state phase is long compared to the set-up phase.

#### 1. *Advertisement Phase*

Initially, when clusters are being created, each node decides whether or not to become a cluster-head for the current round. This decision is based on the suggested percentage of cluster heads for the network (determined a priori) and the number of times the node has been a cluster-head so far. This decision is made by the node n choosing a random number between 0 and 1.If the number is less than a threshold T(n), the node becomes a cluster-head for the current round. The threshold is set as:

$$T(n) = \begin{cases} \frac{P}{1-P*(r \bmod \frac{1}{P})} & \text{if } n \in G \\ 0 & \text{otherwise} \end{cases}$$

Where P = the desired percentage of cluster heads (e.g., P = 0.05), r = the current round, and G is the set of nodes that have not been cluster-heads in the last *1/P* rounds. Using this threshold, each node will be a cluster-head at some point within *1/P* rounds. During round 0 (*r* = 0), each node has a probability *P* of becoming a cluster-head. The nodes that are cluster-heads in round 0 cannot be cluster-heads for the next *1/P* rounds.

Thus the probability that the remaining nodes are cluster-heads must be increased, since there are fewer nodes that are eligible to become cluster-heads. After *1/P* -1 rounds, T=1 for any nodes that have not yet been cluster-heads, and after *1/P* rounds, all nodes are once again eligible to become cluster-heads. Future versions of this work will include an energy-based threshold to account for non-uniform energy nodes. In this case, we are assuming that all nodes begin with the same amount of energy and being a cluster-head removes approximately the same amount of energy for each node.

Each node that has elected itself a cluster head for the current round broadcasts an advertisement message to the rest of the nodes. For this "cluster-head-advertisement" phase, the cluster-heads use a CSMA MAC protocol, and all cluster-heads transmit their advertisement using the same transmit energy.

#### 2. *Cluster Setup Phase*

After each node has decided to which cluster it belongs, it must inform the cluster-head node that it will be a member of the cluster. Each node transmits this information back to the cluster-head again using a CSMAMAC protocol. During this phase, all cluster-head nodes must keep their receivers on.

#### 3. *K-means Clustering*

K-Means [9] Training starts with a single cluster with its center as the mean of the data. This cluster is split into two and the means of the new clusters are iteratively trained.

2. Repeat until no change
3. Assign each node to the cluster of the nearest CH
4. Calculate the mean value of the clusters

*B. Clustering based on Fuzzy Logic*

A fuzzy logic approach to cluster-head election is proposed based on three descriptors - energy, concentration and centrality. Depending upon network configuration a substantial increase in network lifetime can be accomplished as compared to probabilistically selecting the nodes as cluster-heads using only local information. For a cluster, the node elected by the base station is the node having the maximum chance to become the cluster-head using three fuzzy descriptors - node concentration, energy level in each node and node centrality with respect to the entire cluster, minimizing energy consumption for all nodes consequently increasing the lifetime of the network.

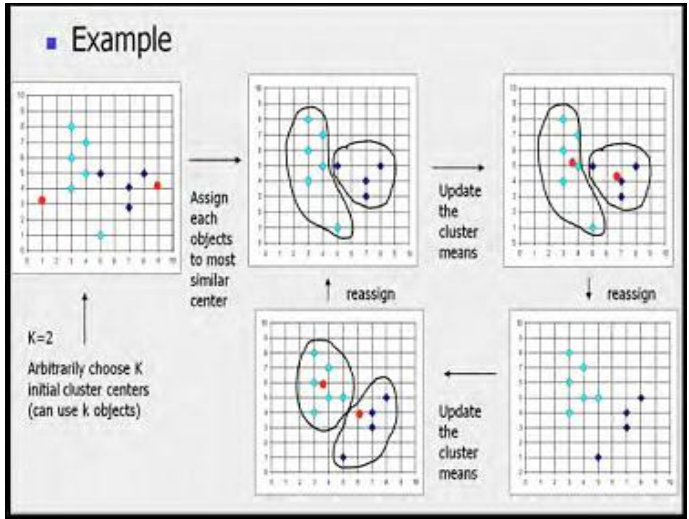

Figure 3. K-Means clustering process.

These two clusters are again split and the process continues until the specified number of clusters is obtained. If the specified number of clusters is not a power of two, then the nearest power of two above the number specified is chosen and then the least important clusters are removed and the remaining clusters are again iteratively trained to get the final clusters. When the user specifies random start the algorithm generates the k cluster centers randomly and goes ahead by fitting the data points in those clusters. This process is repeated for as many random starts as the user specifies and the Best value of start is found. The outputs based on this value are displayed. K Means is an unsupervised clustering algorithm. The data set into k clusters using the cluster mean value. It is iterative in nature. The distance between the nodes is calculated using the Euclidean distance.

The Euclidean distance between two data points, $X_1 = (x_{11}, x_{12}… x_{1n})$ and $X_2 = (x_{21}, x_{22}… x_{2n})$

$$Dist(x_1, x_2) = \sqrt{\left(\sum((x_{1i} - x_{2i})^2)\right)} \quad (2)$$

This distance is used to calculate the distance between all the nodes. This distance helps in determining which nodes will be clustered in which particular cluster. K means clustering is the simplest clustering algorithm. This algorithm makes an assumption that the network is static and homogeneous. There is a drawback of this algorithm that there is a difficulty in finding the center.

**K-Means Clustering Algorithm:**

Steps
1. **Arbitrarily choose k nodes as initial CH having maximum energy**

**Fuzzy Based Clustering Algorithm:**
Input:
   *D = { d1, d2, d3... di... dn }* // Set of n data points.
   k =Number of desired clusters
Output:
   Objects in belongs to more than one groups or class.
Methods
1. **Choose a number of clusters and assign randomly to each point coefficients for being in the clusters.**
2. **Assign each points di to the cluster which has the highest membership values.**
3. **Compute the centroid for each cluster using the below formula.**

$$c_j = \frac{\sum_{i=1}^{n} u_{ij}^m \cdot x_i}{\sum_{i=1}^{n} u_{ij}^m} \quad (3)$$

4. **For each point, compute its membership values of being in the clusters, using the below formula**

$$u_{ij} = \frac{1}{\sum_{k=1}^{c} \left(\frac{\|x_i - c_j\|}{\|x_i - c_k\|}\right)^{\frac{2}{m-1}}} \quad (4)$$

5. **Repeat step 3 to 4 until the algorithm has converged**

Our proposed system makes use of a combination of the concepts of LEACH protocol and EECS method with Fuzzy based clustering Algorithm. The concepts of the fuzzy based clustering are used to grouping the sensor networks and finding the better cluster head, etc.

IV. EXPERIMENTAL RESULTS AND DISCUSSION

A. *Implementation and Simulation* In this section we have mentioned the details about the implementation

of the proposed algorithm and the results found after the implementation. The details are as follows:

*1. Simulation Set up*

We simulated the proposed algorithm in NS 2.29 [7]. We found results for placing the cluster heads with minimum distance separated as well as placing the cluster heads randomly over the grid. We also calculated the intra cluster and inter cluster distance. Analyses the network in terms of packet delivery ration, Energy consumption for transmission, dropped packets and found that the network works well.
For the simulation experiments, following parameters were used:

> Tx Antenna Gain Gt = Rx
> Antenna Gain Gr=1
> Antenna Height (Ht) =1.5m,
> Base Station Location was (500,200)

*2. Simulation Results*

As per mentioned in [8], 5% of total number of cluster gives the better performance in the network. We have clustered the network in same number of clusters. We have initiate the intra cluster distance and inter cluster distance of the cluster. Results have shown that, we have mentioned that the cluster heads can be placed randomly or separated with some minimum distance. Results show that if the cluster heads are separated with some minimum distance it gives the better performance. We have considered the minimum distance as 50 meters.

Table 1. Simulation Parameters

| S.no | No. Item Description Parameter | No. Item Description Parameter |
|---|---|---|
| 1 | Simulation Area | 1000x 1000 |
| 2 | No. of Nodes | 100 |
| 3 | Radio Propagation Model | Two ray ground |
| 4 | Channel Type | Channel/ Wireless channel |
| 5 | Antenna Model | Antenna/Omniantenna |
| 6 | Interface Queue Type | Queue/Drop Tail/PriQueue |
| 7 | Link Layer Type | LL |
| 8 | Energy Model | Battery |
| 9 | Min Packets in ifq | 30 |

*3. Execution of clustering schemes*

The execution of a clustering algorithm can be supported out at a centralized authority or in a distributed way at local nodes. Centralized approaches require global. The performance of the schemes is evaluated considering network lifetime as a parameter which is defined as the time until the last node dies in the network. Network lifetime is measured using two different yard-sticks:

- **a. Number of nodes alive in the network** - more number of nodes alive implies network lifetime lasts longer.
- **b. Number of messages received at BS** - more number of messages received at BS denotes more number of nodes is alive in the network leading to longer network lifetime.

*4. Network performance analysis*

To validate the performance of LEACH, EECS and HEED Clustering for our experiments, we used a 100 node network where nodes were randomly distributed between(x=0, y=0) and (x=100, y=100) with the BS at location(x=50, y=175). The bandwidth of the channel was set to 1 Mb/s, each data message was 500 bytes long, and the packet header for each type of packet was 25 bytes long.

The number of nodes alive in over time for different method is obtained and listed in the below Table2.

Table 2. Number of nodes alive in over time

| S. No | Number of nodes alive over time. (In sec) | Number of nodes alive | | |
|---|---|---|---|---|
| | | LEACH | HDDP | EECS |
| 1 | 100 | 100 | 100 | 100 |
| 2 | 200 | 100 | 100 | 100 |
| 3 | 300 | 88 | 96 | 100 |
| 4 | 400 | 75 | 86 | 93 |
| 5 | 500 | 40 | 51 | 75 |
| 6 | 600 | 23 | 30 | 64 |
| 7 | 700 | 8 | 15 | 31 |
| 8 | 800 | 0 | 3 | 7 |
| 9 | 900 | 0 | 0 | 0 |
| 10 | 1000 | 0 | 0 | 0 |

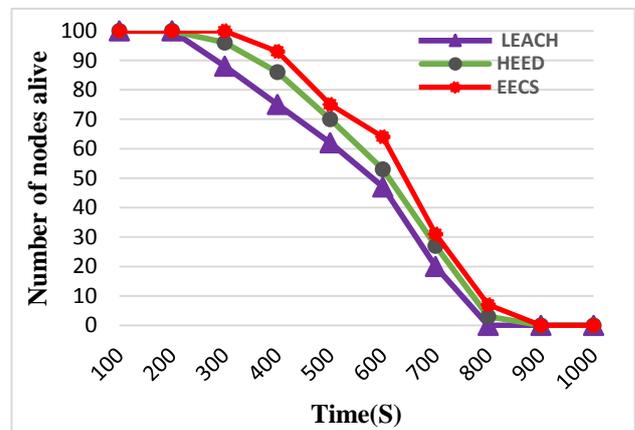

Figure 4. Comparison chart for Total number of alive nodes in the LEACH, HEED, EECS

The improvement increased through EEPSC compared to LEACH and HEED is further showed in Figure 4, which specifies the lifetime of network is extended and the overall number of messages received at base station is increased. With LEACH and HEED, all nodes remain alive for 245and 270 seconds before the first node dies, while in EECS, all nodes remain alive for 360 seconds, which is 39% more than LEACH and HEED. Figure 10 clearly indicate the advantages of EEPSC over LEACH and HEED in terms of increasing network lifetime.

The number of live nodes in the system decreases to about less than 5 nodes at time 10000, but the network is still well connected and only the nodes' redundancy is removed. From this time, the nodes die quickly, so the connectivity of the network and its coverage rapidly decrease. Since the data rate of EECS is larger than LEACH, HEED, the deterioration is steeper.

## 5. Messages received at Base station

The total number of messages received at base station with three different methods LEACH, HEED and EECS are obtained and depicted in the below Table 3.

Table 3. Message received at Base Station (BS)

| S. No | Time (in Sec) | Number of Messages received at BS | | |
|---|---|---|---|---|
| | | LEACH | HEED | EECS |
| 1 | 70 | 3005 | 5487 | 6647 |
| 2 | 140 | 5741 | 7845 | 9974 |
| 3 | 210 | 9561 | 16578 | 22478 |
| 4 | 280 | 16245 | 30458 | 40578 |
| 5 | 350 | 23054 | 37845 | 55174 |
| 6 | 420 | 29595 | 44578 | 59428 |
| 7 | 490 | 34289 | 53541 | 64825 |
| 8 | 560 | 39648 | 59864 | 67845 |
| 9 | 630 | 46254 | 60247 | 70458 |
| 10 | 700 | 51540 | 60564 | 72894 |
| 11 | 770 | 55800 | 63584 | 74589 |
| 12 | 840 | 56250 | 64875 | 75415 |
| 13 | 910 | 56252 | 66455 | 75412 |
| 14 | 980 | 56250 | 66453 | 75415 |
| 15 | 1000 | 56250 | 66458 | 75415 |

From the above figure 5, it clearly shows that the overall number of messages received at base station is increased in EECS method for all different timeline, the EECS obtain the better performance than LEACH and HEED method.

## 6. Clustering performance analysis

In this section the two different clustering methods are involved and compared in the process of time taken for first node to die in WSN, the results are obtained from the two different methods and listed in the below Table4.

Table 4. Times taken for first node to die in WSN

| S. No | Rounds | Time taken for first node dies | |
|---|---|---|---|
| | | Kmeans | Fuzzy Method |
| 1 | R1 | 1584 | 1348 |
| 2 | R2 | 1862 | 1574 |
| 3 | R3 | 2075 | 2104 |
| 4 | R4 | 1727 | 1384 |
| 5 | R5 | 1973 | 1754 |
| 6 | R6 | 2485 | 2754 |
| 7 | R7 | 2155 | 2014 |
| 8 | R8 | 1687 | 1548 |
| 9 | R9 | 2457 | 2105 |
| 10 | R10 | 1824 | 1687 |

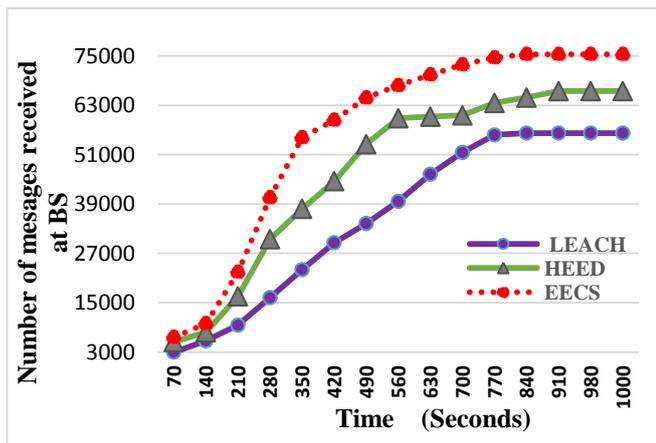

Figure 5. Comparison chart for message received at Base Station (BS)

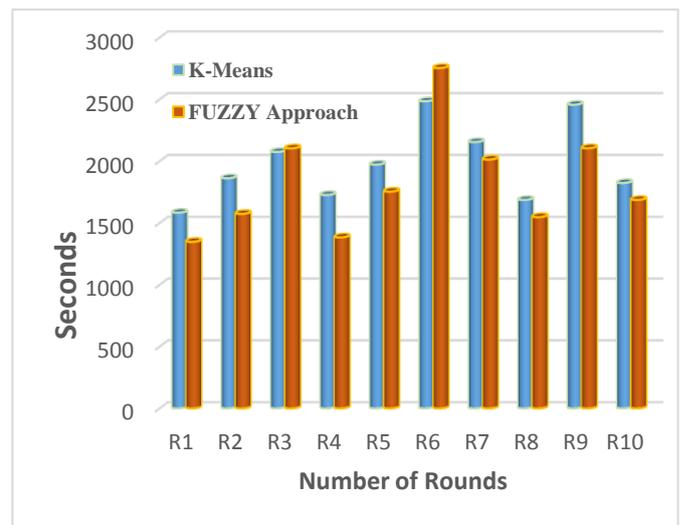

Figure 6. Comparison chart of Time taken for first node to die in WSN

As seen from **Table 4,** the times taken for first node to die are comparable in the case of the fuzzy logic approach and the Kmeans approach. As seen from Figure 6, the fuzzy approach leads to the time steps after which the first node dies to be much later than that of Kmeans method. Also all the nodes die almost at the same time as opposed to the random fashion in which nodes die as in the case of Kmeans method. The death of the last node in Kmeans occurs much later than that in the fuzzy logic approach. Therefore a clustering algorithm allows the system to work for a longer time although the performance of the system may reduce. Whereas in case of fuzzy logic approach the system gives the maximum performance till the end and dies instantly.

### 7. *Iteration level analysis*

Iteration level is defined that the number of executions required to converge the clustering process. Different Clustering algorithms are compared for their performances using the time required to cluster the nodes in wireless sensor network. The execution time is varying while selecting the number of initial cluster centroids. The execution time is increased and decreased when the number of cluster head is increased. The obtained results are depicted in the following Table 5.

Table 5. Execution level for Kmeans and Fuzzy based clustering method.

| S. No | Cluster Head | Number of iterations | |
|---|---|---|---|
| | | K-Means | Fuzzy Method |
| 1 | 10 | 15 | 13 |
| 2 | 20 | 10 | 6 |
| 3 | 30 | 22 | 13 |
| 4 | 40 | 18 | 10 |
| 5 | 50 | 13 | 7 |
| 6 | 60 | 9 | 13 |
| 7 | 70 | 15 | 10 |
| 8 | 80 | 10 | 8 |
| 9 | 90 | 4 | 7 |
| 10 | 100 | 9 | 5 |

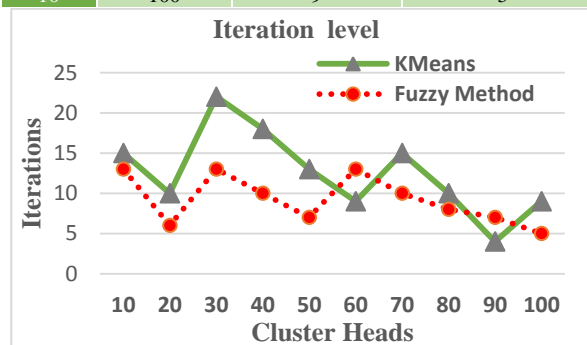

Figure 7. Iteration level chart for kmeans and Fuzzy based clustering methods

From the above figure 7, it clearly show that the fuzzy based clustering algorithm is executed very faster than kmeans clustering methods. In fuzzy based clustering method the intra distance between the cluster head and sensor nodes is too small than Kmeans clustering method. In fuzzy based clustering method, the clustering process is converged with minimum number of iterations than kmeans clustering algorithm for most of the different clustering heads. Thus the computational complexities required in the Fuzzy based clustering is much lesser than Kmeans clustering method. Hence the fuzzy based clustering methods achieve better performance than kmeans clustering methods for node clustering on wireless sensor network.

### V. CONCLUSION

Wireless sensor network is a current area for research now days, due to vast potential usage of sensor networks in different areas. A sensor network is a comprised of sensing, processing, communication ability which helps to observe, instrument, react to events and phenomena in a specified environment. Clustering is a useful topology-management approach to decrease the communication overhead and adventure data aggregation in sensor networks. We have classified the different clustering approaches according to the clustering criteria and the entity responsible for carrying out the clustering process. Our thesis work included the study of wireless sensor network, clustering, cluster head selection and other energy efficient communication protocols for WSN, since it was earlier proposed that clustering method improves the network lifetime.

We have studied and implement the three different cluster head selection methods LEACH, HEED and EECS which is compared the performance of each of the clustering methods. It was found that EECS give a much reduced network lifetime as compared to LEACH and HEED. The experimental results shows that the EECS with Fuzzy based clustering method received more number of messages at Base Station (BS) than LEACH and HEED. However the proposed Fuzzy based clustering method along with the EECS method of cluster head selection provides a much increased performance with a faster convergence as compared to other techniques. In clustering process the Fuzzy based clustering methods is better than kmeans method due to the clustering process is converged with minimum iteration in Fuzzy clustering. Our algorithm tries to change the cluster head of the nodes if the CH is running out of the energy, it helps to minimize the dropped packets.

Different types of cluster head selection methods and different clustering methods are used to improve the network life time, messages received at base station and its performance in wireless sensor network is our future work.

**REFERENCE**

[1]  Heinzelman W, Chandrakasan A, Balakrishnan H. Energy Efficient Communication Protocol for Wireless Micro sensor Networks Proc. 33rd HICSS, 2000.

[2]  Heinzelman W. R, A. Chandrakasan, and H.Balkrishnan, "Energy-Efficient Communication Protocol for Wireless Micro sensor Networks", in Proceedings of 33rd Hawaii International Conference on System Science, Vol. 2, Jan. 2000, pp.1-10.

[3]  Inbo Sim, KoungJin Choi, KoungJin Kwon and Jaiyong Lee, Energy Efficient Cluster header Selection Algorithm in WSN International Conference on Complex, Intelligent and Software Intensive Systems, 978-0-7695-3575-3/09.

[4]  Lindsey S., C.S. Raghavendra, PEGASIS: power efficient gathering in sensor information systems, in: Proceedings of the IEEE Aerospace Conference, Big Sky, Montana, March 2002.

[5]  Lotf J.J., S.H.Hosseini Nazhad Ghazani," Clustering of Wireless Sensor Networks Using Hybrid Algorithm", Australian Journal of Basic and Applied Sciences, 5(8): 1483-1489, 2011.

[6]  Mao Y., L. Chengfa, C. Guihai, and J. Wu "EECS: An energy efficient clustering scheme in wireless sensor networks". In *Proceedings IPCCC, IEEE 24th International*, 535–540

[7]  NS2, Network Simulator, World Wide Web: http://www.isi.edu./nsnam/ns/nsbuild.html 2004.

[8]  Rajesh Patel, Sunil Pariyani, Vijay Ukani, Energy and Throughput Analysis of Hierarchical Routing Protocol (LEACH) for Wireless Sensor Network IJCA (0975 8887) Volume 20 No.4, April 2011.

[9]  Sauravjoyti Sarmah and Dhruba K. Bhattacharyya. "An Effective Technique for Clustering Incremental Gene Expression data", IJCSI International Journal of Computer Science Issues, Vol. 7, Issue 3, May 2010.

[10] Younis O. and S. Fahmy, "HEED: A Hybrid Energy-Efficient Distributed Clustering Approach for Ad Hoc Sensor Networks," IEEE Transactions on Mobile Computing, vol. 3, no. 4, Oct-Dec 2004.